\documentclass[letterpaper, 10 pt, conference]{ieeeconf}  

\IEEEoverridecommandlockouts                              

\usepackage{times} 
\usepackage{amsmath} 
\usepackage{amssymb}  

\usepackage[acronym]{glossaries}
\usepackage{cite}
\usepackage{caption}
\usepackage{subcaption}
\usepackage{graphicx}
\usepackage{url}

\title{\LARGE \bf
A Framework for Operational Volume Generation for Urban Air Mobility Strategic Deconfliction
}

\author{Ellis L. Thompson$^{1}$ and Yan Xu$^{2}$ and Peng Wei$^{3}$
\thanks{$^{1}$Ellis L. Thompson is with the School of Engineering and Applied Science,
        George Washington University, Washington, DC 20052, USA
        {\tt\small thompson\_e@gwu.edu}}%
\thanks{$^{2}$Yan Xu is with the Centre for Autonomous and Cyberphysical Systems, Cranfield University, Bedford, MK43 0AL, UK
        {\tt\small yanxu@cranfield.ac.uk}}%
\thanks{$^{3}$Peng Wei is with the School of Engineering and Applied Science,
        George Washington University, Washington, DC 20052, USA
        {\tt\small pwei@gwu.edu}}%
}

\newacronym{atm}{ATM}{Air Traffic Management}
\newacronym{conops}{ConOps}{Concept of Operations}
\newacronym{evtol}{eVTOL}{electric Vertical Take-Off and Landing}
\newacronym{easa}{EASA}{European Aviation Safety Authority}
\newacronym{faa}{FAA}{Federal Aviation Administration}
\newacronym{fmt}{FMT}{Fast Marching Tree}
\newacronym{nfz}{NFZ}{No-Fly Zone}
\newacronym{ov}{OV}{Operational Volume}
\newacronym{prm}{PRM}{Probabilistic Roadmap}
\newacronym{rrt}{RRT}{rapidly-exploring random tree}
\newacronym{rrtfn}{RRT*FN}{rapidly-exploring random tree fixed nodes}
\newacronym{rrtfnd}{RRT*FND}{rapidly-exploring random tree fixed nodes - dynamic}
\newacronym{uam}{UAM}{Urban Air Mobility}
\newacronym{uas}{UAS}{Unmanned Aircraft System}
\newacronym{uav}{UAV}{Unmanned Aerial Vehicle}
\newacronym{2d}{2D}{two-dimensional}
\newacronym{3d}{3D}{three-dimensional}
\newacronym{4d}{4D}{four-dimensional}

\begin{document}

\maketitle
\thispagestyle{empty}
\pagestyle{empty}

\begin{abstract}
Strategic pre-flight functions focus on the planning and deconfliction of routes for aircraft systems. The urban air mobility concept calls for higher levels of autonomy with onboard and en route functions but also strategic and pre-flight systems. Existing endeavours into strategic pre-flight functions focus on improving the route generation and strategic deconfliction of these routes. Introduced with the urban air mobility concept is the premise of operational volumes. These 4D regions of airspace, describe the intended operational region for an aircraft for finite time. Chaining these together forms a contract of finite operational volumes over a given route. It is no longer enough to only deconflict routes within the airspace, but to now consider these 4D operational volumes. To provide an effective all-in-one approach, we propose a novel framework for generating routes and accompanying contracts of operational volumes, along with deconfliction focused around 4D operational volumes. Experimental results show efficiency of operational volume generation utilising reachability analysis and demonstrate sufficient success in deconfliction of operational volumes.
\end{abstract}


\section{INTRODUCTION}
\label{sec: introduction}

\gls{uam} concept is generally described in \cite{UAMvision, easaUAM, faauam} as: a safe and efficient, highly automated, aviation transportation infrastructure, for passengers and cargo, operating at low altitudes. The highly automated nature of this concept favours emerging technologies for aircraft, such as \glspl{uas} and \gls{evtol} vehicles, as well as \gls{atm} and route planning systems. The \gls{easa} predicts \glspl{uas} to become prevalent in European cities by 2025\cite{easaUAM}, and the \gls{faa} estimates 828 thousand unmanned commercial aircraft by 2024\cite{faaforecast}.

The prediction of high-density \gls{uam} airspace brings with it the requirement for both highly automized aircraft and \gls{atm} techniques. Introducing autonomy at the strategic pre-flight, level can provide efficient route planning and deconfliction for aircraft wishing to enter the airspace. Additionally, autonomous aircraft and \gls{atm} systems can provide tactical instructions to ensure the airspace remains safe with efficient traffic flow. 

A \gls{uam} operator is introduced as an entity responsible for the management of \gls{uam} specific operations \cite{FAAConOpsV1}. A \gls{uam} service provider therefore is any entity offering services to assist \gls{uam} operations. One such role that either an operator or service provider is responsible for is route planning and sufficient deconfliction for safe and efficient flight. The concept of \gls{4d} regions of airspace, including time, is presented as \glspl{ov} \cite{FAAConOps,SESARCORUS}, with the intention that those operating aircraft remain within their designated \glspl{ov} for the duration of the flight. This consequently becomes an agreed upon \textit{contract} between the aircraft operator and airspace, of where the aircraft is allowed to operate and for what duration.

In this paper we propose a framework for generating deconflicted routes as well as \glspl{4d} \glspl{ov} over the routes. Motivated by the work presented in SkyTrakx \cite{skytrakx} we introduce route planning and deconfliction to provide initial candidate routes for the \glspl{ov} to be generated over. We additionally expand on their definition of an \gls{ov}, resolving assumptions around \gls{ov} boundary arrival times which could potentially cause inaccuracies in the temporal element of \glspl{ov}. We further expand on the definition of an \gls{ov} given in \cite{skytrakx} by providing a definition that is sufficiently more adaptive to an uncertain airspace environment, by including the distribution of simulated aircraft in our modified definition. To build the \glspl{ov} our approach leverages the reachability analysis framework identified in \cite{dryvr}. To generate the routes for \gls{ov} generation we propose a \gls{rrt} based algorithm focusing on the cruise phase of flight.

The remainder of this document is formatted as following: Section \ref{sec: related} covers related works, we then formally define both a \textit{contract} and \textit{operational volume} and provide our implementation for generating \glspl{ov} in Section \ref{sec: contracts and ovs}. The approach used for route generation is presented in Section \ref{sec:Route Generation}. Experimental results are covered in Section \ref{sec:experiments}, followed by a discussion on the proposed framework in Section \ref{sec: discussion}. Finally, concluding remarks are provided in Section \ref{sec:conclusion}.

\section{RELATED WORKS}
\label{sec: related}

\subsection{Concept of Operations}
Stakeholders in the field of \gls{uam}, including the \gls{faa}, NASA and Europe's SESAR Joint undertaking, provide outlines for the conceptual \gls{uam} requirements. Provided in the technical documents \cite{FAAConOps, SESARCORUS, catapultutm, sdNASA}, are outlines into the safe operating standards expected to be imposed in \gls{uam} airspace. In addition to the U-Space operating regions and requirements, such as defining maximum altitudes, one key area referenced heavily is the imposition of operating volumes unique to each flight within the U-Space. Defined as reserved regions of \gls{4d} airspace and often depicted as either linear monotonically progressing tubes or polygons, there is a heavy implication that, within the conceptual \gls{uam} U-Space, operators will be required to define their expected operating area pre-flight.

\subsection{Route Planning}
Route planning algorithms provide potential candidate routes from an origin to a destination. Firstly, shortest path algorithms, such as Dijkstra's algorithm, A* and D*\cite{Dijkstra1959,Hart1968,351061}, generate the shortest or, in the case of A* and D*, the minimum cost path between two nodes. Used heavily in the field of Computer Science, A* and D*, work to optimise some heuristic function, commonly distance, always expanding the node with the lowest heuristic cost. The route can be considered to be guaranteed optimal if an admissible heuristic is found, that is a heuristic function that never overestimates the cost of the shortest path. The inverse of this would be a non-admissible heuristic, where the optimality of the generated path cannot be guaranteed but in many cases a near-optimal path can be found in less time.

A second class of route planning algorithms can be derived from motion control in robotics. Included in this classification are, \gls{rrt} based approaches as well as \glspl{prm} and \glspl{fmt} \cite{rrt,rrtfn,rrtfnd,rrtrope,508439,janson2015fast}. These approaches are often not optimal and largely work on the principal of paths between randomly generated points, meaning the solution paths are rarely reproducible. However, they have the benefit of, in less complex environments, regularly finding a path in less time than A* and D*.

\subsection{Strategic Deconfliction}
Strategic deconfliction\cite{Sacharny2022}, is specifically concerned with pre-flight stratagem to provide efficient use of the airspace while reducing the likelihood of en route conflicts. Encompassed in this is the act of route planning and re-planning with a greater emphasis on re-planning for deconfliction. Most strategic deconfliction approaches \cite{chaimatanan:hal-00868450,Tang2016,10.1007/978-3-319-42902-1_38,bertram2020scalable} predict the aircraft's position at a given time and compares that with the predicted position of other airspace users. If a conflict, or conflicts, are detected then one of two approaches are usually employed: as in \cite{Tang2016,10.1007/978-3-319-42902-1_38} delays can be imposed to the schedule proposing the aircraft departs later than its previously scheduled time. Alternatively, as in \cite{chaimatanan:hal-00868450,Tang2016} re-planning can occur to change the original filed route proposing either changes to segments or the entire flight plan.

Intuitively, it is often favoured to propose a new departure time over re-planning the route due to its simpler nature and implementation for an already existing route. The latter however can be useful if no prior route exists, that is, the route is strategically deconflicted at the time of generation.  

\subsection{Reachability Analysis}
As mentioned, our proposed framework uses reachability analysis to build the \glspl{ov} from simulation data. Reachability analysis is the process of computing reachable sets given some initial conditions usually accompanied by either simulation data or real data. DryVR\cite{dryvr} provides data-driven approach for verification of automotive systems. In this approach reachable sets are generated by simulation and accompanying reach tubes, conservative volumes of reachable states over time, are generated. The reach tubes are generated through learning a discrepancy function and the related sensitivities of the simulation-generated data in relation to the initial given state.

An alternative methods, identified in \cite{ctrl,Gruenbacher_2020,gotube}, propose approaches utilising Lagrangian-based methods, with \cite{Gruenbacher_2020,gotube} providing an analytical approach to reach tube generation. The benefit of these methods over the globally learned discrepancy function of DryVR is that, unlike DryVR, these methods are less susceptible to the wrapping effect\cite{Neumaier1993}, overall producing tighter volumes for the reach tubes. 

\subsection{Operational Volumes, Geofences and SkyTrakx}
SkyTrakx\cite{skytrakx} proposes an approach to generating a contract of \glspl{ov} for a route and specific aircraft type by performing a reachability analysis over a finite time horizon. The SkyTrakx framework demonstrates the application of reachability-based \gls{ov} generation, for light \glspl{uas}, for both rotor and fixed-wing aircraft. They further allow for two generation types of \glspl{ov}: conservative and aggressive, with the latter producing tighter bounded \glspl{ov}. However, assumptions are made about the arrival time of aircraft at \gls{ov} boundaries, potentially causing the temporal element of the \glspl{ov} to become inaccurate. 

Dynamic geofences are also proposed in \cite{doi:10.2514/6.2016-3453} for coordination of low-altitude \glspl{uas}. This approach uses knows of the aircraft dynamics to build a \gls{3d} volumes of airspace, referred to as a geofences, that the aircraft must visit in sequence. Unlike \cite{dryvr} the volumes generated here are conservative to the extremes of the aircraft performance. It could be expected however, that this approach is susceptible to inaccuracies, significantly so as there is little provision added for the complexity introduced by wind. Yet, \cite{doi:10.2514/6.2016-3453} has considered the possibility for delays producing a slightly less stringent requirement, or rather assumption, that the arrival time at volume boundaries can be guaranteed, creating a possibly more adaptable approach.

In \cite{app12020576}, \textit{keep-in} geofences are generated in tandem with a route generation. Monte Carlo simulations were used to propose volumisation strategies. An overlap between volumes was proposed as part of the strategy to account for the \gls{uas} flying faster or slower than expected, this effectively creates a buffer region where the aircraft can inhabit the most suitable volume for a given time. For the route planning and deconfliction, the flight path was optimised against flight time, distance and energy usage. Additionally, the approach maintained a minimum buffer, defined by the user, around both buildings and other aircraft through the use of obstacle-free visibility graphs. 

\section{CONTRACTS AND OPERATIONAL VOLUMES}
\label{sec: contracts and ovs}

\subsection{Defining Contracts and Operational Volumes}
\label{sec:Defining Contracts and OVs}
Before we can define what is meant by the term \textit{contract} we must first provide a definition of an \gls{ov}. We can initially describe an \gls{ov} as: A \gls{4d} region of airspace that describes both the positional operational area and the duration over which the volume is considered active. As such, if we represent the airspace as $\mathcal{X} \subseteq \mathbb{R}^3$, then a single \gls{ov} is represented as tuple in Equation \ref{eq: initial ov}.
\begin{equation}
    \mathcal{C} = (R_1,T_1),(R_2,T_2),...,(R_k,T_k)
    \label{eq: initial ov}
\end{equation}
Where, $R_n \subseteq \mathcal{X}$ is a subset volume of the airspace and $T_n$ is monotonically increasing time. This initial definition is derived in \cite{skytrakx}, and provides the foundation for the expansion we will apply. It should also be noted that this definition of an \gls{ov} consists of several smaller volumes over discrete time intervals, the reason for this will be explained further in this section. At present though we are able to define the total area of this finite time operational volume as Equation \ref{eq:total_airspace} and the total duration as $\mathcal{C}^{dur} = T_k-T_1$.
\begin{equation}
     \mathcal{C}^{vol} = \bigcup_{n=1}^{k}R_n
    \label{eq:total_airspace}
\end{equation}

The definition of a contract ($\Sigma$) can then be determined as: The sequence of \glspl{ov} over a given proposed trajectory for a single aircraft:
\begin{equation*}
    \Sigma = (\mathcal{C}_1,\mathcal{C}_2,...,\mathcal{C}_i)
\end{equation*}

Additionally, we can now impose two further requirements into the definition of an \gls{ov}: (1) For any pair of \glspl{ov}, $\forall(\mathcal{C}_i, \mathcal{C}_j) \in \Sigma$, then $\mathcal{C}_i^{dur} = \mathcal{C}_j^{dur}$; (2) By extension the interval at which the volumes in $\mathcal{C}_i$ are generated must be regular both within $\mathcal{C}_i$ and $\forall\mathcal{C}_n \in \Sigma$.

It is at this point our definition begins to deviate from that given in \cite{skytrakx}. Where the assumption is made that, for two successive \glspl{ov}, $\mathcal{C}_i$ and $\mathcal{C}_j$ the following is true: $\mathcal{C}_i(T_k) \equiv \mathcal{C}_j(T_1)$, that is there exists a hard boundary between \glspl{ov} such that they do not overlap temporally. In our system we introduce the constant $\delta$ such that now $\mathcal{C}_j(T_1) = \mathcal{C}_i(T_{k-\delta})$ where $\delta \in [0,\mathcal{C}^{dur})$.

We are now able to make one more change to the definition of an \gls{ov}. For deconfliction of \glspl{ov} some distribution of aircraft position must be known at a given time. As such we can modify the definition given in Equation \ref{eq: initial ov}, like so:
\begin{equation}
    \mathcal{C} = (R_1,T_1,D_1),(R_2,T_2,D_2),...,(R_k,T_k,D_k)
    \label{eq: ov definition}
\end{equation}

Here, we have introduced the value $D_n$, representing the distribution of aircraft positional data at each discrete time interval. Defined more concretely later, this value now allows for the probabilistic evaluation of the aircraft position at any given time within the bounds of an \gls{ov} or contract.

\subsection{Generating Operational Volumes}
\label{sec:ov generation}
For a given, predefined route, \glspl{ov} are generated using an adaptation of the DryVR\cite{dryvr} framework. Included within the DryVR framework is: ``a probabilistic algorithm for learning sensitivity of the continuous trajectories from simulation data'' as well as an algorithm that then uses these sensitivities to perform a reachability analysis and generate a reach tube. We leverage these algorithms by utilising simulation data obtained with the BlueSky\cite{bluesky} \gls{atm} simulator to build a contract of \gls{4d} \glspl{ov} over a predefined trajectory.

In the generation of the \glspl{ov} two factors need to be considered: the first, as outlined previously, \glspl{ov} exist for a predetermined duration $\mathcal{C}^{dur}$ but, for a \gls{ov} $\mathcal{C}_n, n>0$, the start time is offset from by some fixed amount $\delta$ from the previous \gls{ov}. The second factor, not inherently obvious, is the inclusion of uncertainty. While we are able to simulate the aircraft's trajectory initial factors, as well as variations in speed, acceleration, climb rates and wind are not known. As such a level of reasonable uncertainty needs to be maintained when running these simulations.

The generation of the simulated trajectory data and accompanying \glspl{ov} are performed in three stages: (1) Uncertainty is introduced into the simulation during the initialisation of aircraft starting states; (2) The simulation is run, for some duration ($t_d = \mathcal{C}^{dur}$) with aircraft states being stored are regular intervals, for example 1 second; (3) The DryVR framework is used on a sample of output trajectories and then verified on the remaining trajectories to build the \gls{ov} bounds. This three-step approach is repeated until the aircraft have reached the terminal waypoint in the simulation. The resultant \glspl{ov} are then combined into a contract.

\subsubsection{Initialising aircraft states with uncertainty}
\label{sec: ac initilization}
To introduce variation and uncertainty into the simulation $N$ aircraft, between 500 and 2000 are created. For the first initialisation, i.e., from the origin, the position, altitude, heading and speeds are chosen at random from within predetermined bounds. 

After each simulation run to $t_d$, $N$ new aircraft are initialised in the system. Unlike the initial run however, these positions are sampled from the previous batch of trajectories, more specifically, they are sampled from the normal distribution of states from the offset interval $\delta \pm c$, where $c$ is some time constant in seconds. This approach favours generating states closer to the existing \textit{mean state} but provides uncertainty and variation. Additionally, to avoid potentially infeasible states from occurring, the samples are bound to lie either within the previous \gls{ov} or no further than some fixed distance from the previous \gls{ov}.

\subsubsection{Collecting trajectory data from the simulation}
\label{sec: trajectories from simulation}
From the initialised states the BlueSky simulation in run for the duration $t_d$. The step-size and logging interval are independent of each other and in our environment are set as $0.1$ seconds and $1$ second respectively. Two aircraft were used in our simulations: an Amazon octocopter available through the OpenAP\cite{openap} suite and the Airbus Vahana \gls{evtol} vehicle using performance data available in \cite{airbusvahana, pradeep2019phd}. These aircraft provide two different examples of \gls{uam} space users to run evaluations on, although, they are not used together when generating \glspl{ov}.

For the $N$ aircraft in the environment, $N$ states were generated each second with each state storing: record time, latitude, longitude, altitude, heading, vertical speed (m/s), true airspeed (m/s). Noise was also added to the values in the state to further inject uncertainty, mimicking the effects of wind and uncertainty in GPS position.

The result of this second stage is the set of trajectories $S^N$ for $N$ aircraft taken at 1 second intervals from the simulation such that the total duration of the trajectories is equal to $t_d$.

\subsubsection{Generating \glspl{ov} from trajectories}
The resultant trajectories ($S^N$) from the simulation are then used in the DryVR framework to perform reachability analysis and obtain a resultant reach tube. These reach tubes, along with the aircraft distribution data to satisfy Equation \ref{eq: ov definition}, form the set of \glspl{ov} over the prescribed trajectory which form a contract.

The reachability analysis is performed in \gls{3d} space over the current and next states for each time step. Also, for all trajectories only the: record time, latitude, longitude and altitude are provided to the framework. Additionally, to avoid over or under estimation for trajectories with a small geographical footprint, the latitude, longitude and altitude are normalised to values in the range $[0,1]$.

The DryVR framework is provided with a subset of 15-20 trajectories $S^X \subset S^N$. Additionally, the centre trajectory ($S^X_{centre}$) is provided and is used by DryVR to generate the sensitivities. The remaining set ($S^N - S^X$) is used for verification of the \gls{ov} to ensure accuracy, or that the generated volume includes a sufficient number of points from the larger trajectory sample. If the verification is not satisfied then the route and accompanying \glspl{ov} are not released and a re-plan with new parameters is required.

\begin{figure}[ht]
    \centering
    \includegraphics[width=\linewidth]{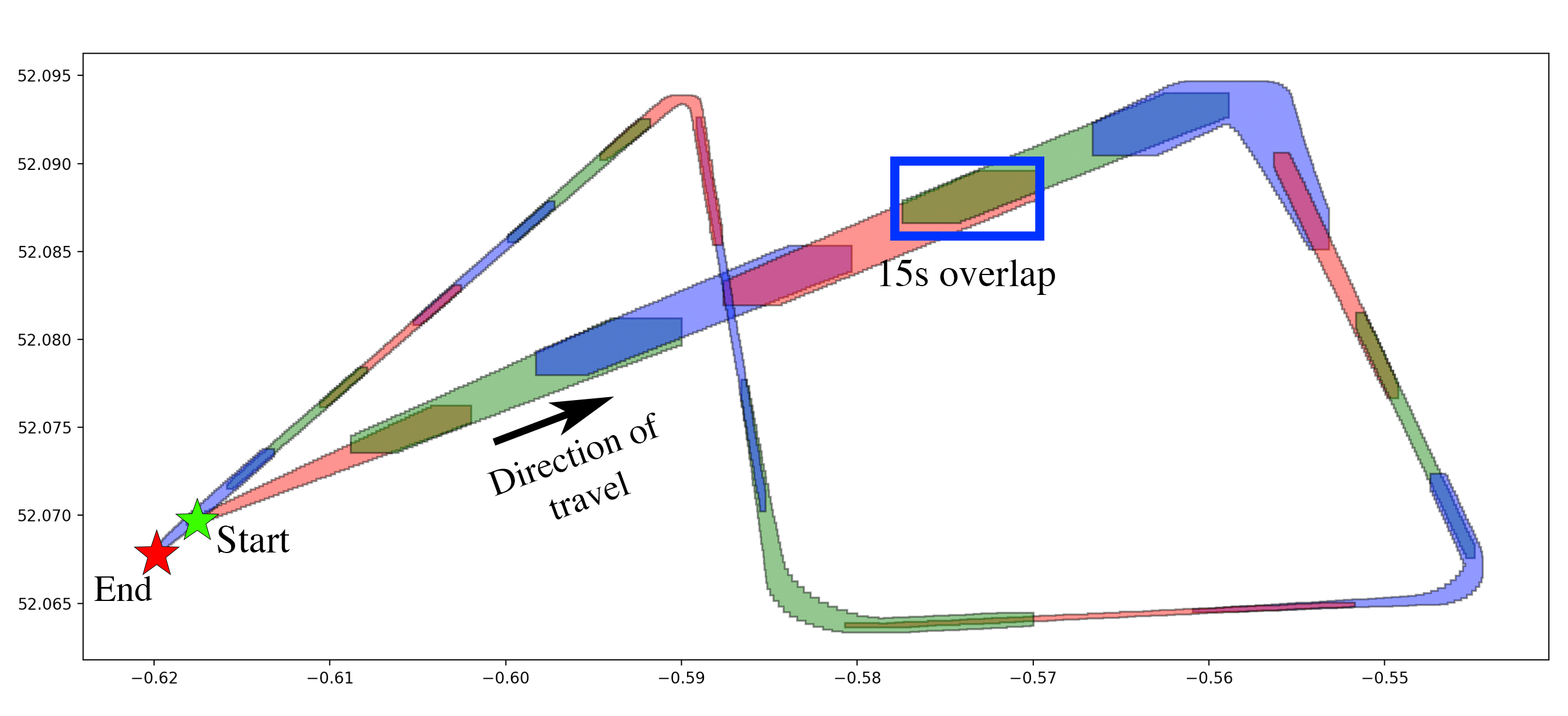}
    \caption{A two-dimensional, top-down view of the 18 \glspl{ov} in a contract. The duration ($t_d$) of each \gls{ov} is 60 seconds and a 15 second offset has been used. Each \gls{ov} has been given a color (red, green or blue) to easily distinguish them.}
    \label{fig:top-down ovs}
\end{figure}

The output of the framework is a reach tube which, combined with the duration $t_d$ and initialisation time ($t_0$), create a single \gls{ov}. When this process is repeated, the \glspl{ov} can then be combined into a contract spanning the proposed flight path as seen in Figure \ref{fig:top-down ovs}.

\subsection{Distribution of Aircraft within \glspl{ov}}
As already emphasised, aircraft are expected to make constant progression along their route, derived from the finite time than an \gls{ov} can exist. To ensure efficient use of the airspace when generating routes, the position of an aircraft within an \gls{ov} can be represented as a distribution and then utilised in route planning to avoid regions of airspace with high likelihoods of aircraft existing.

Included in the definition of an \gls{ov} in Equation \ref{eq: ov definition}, the distribution of aircraft at any time interval within the duration of an \gls{ov} is given as $D_n$. Naturally, if we were to sample any given specific position $s \in \mathcal{C}(t)$ we would not expect to observe an aircraft presence due to the specific precision required. Instead, a grid is overlaid over the area of $\mathcal{C}$ creating a finite number of cells to sample the aircraft distribution over. We can now represent the cells of the grid over the \gls{ov} as $\mathcal{C}_{i,j}$.

So for a specific query position $s$ at time $t$ within the duration of $\mathcal{C}$, and where the function $f(s)\mapsto(i,j)$ returns the relevant pair $(i,j)$ corresponding to a cell coordinate in $\mathcal{C}$. We can now obtain the probability of an aircraft existing for the cell that $s$ belongs to through using Equation \ref{eq: prob}, where $N$ is the total number of aircraft in the simulation and $a$ is a single aircraft.
\begin{equation}
    P(s,t,\mathcal{C})=\frac{1}{N} \left( \sum a \in \mathcal{C}_{f(s)}(t)\right)
    \label{eq: prob}
\end{equation}

\section{ROUTE GENERATION}
\label{sec:Route Generation}

\subsection{An Overview of RRT*FND and RRT-Rope}
For the route generation and adaptation of the \gls{rrt} algorithm was chosen for the significantly faster execution speed in complex environments over algorithms like A*. The method leveraged uses a combination of: \gls{rrtfnd} \cite{rrtfnd} for the initial candidate route generation, and its dynamic property allowing it to reconnect or rebuild disconnected trees, as well as \gls{rrt}-Rope \cite{rrtrope}, which shortens and straightens out the winding routes generated by traditional \gls{rrt} algorithms.

\subsubsection{RRT*FND}
Built on the \textit{fixed node} concept of \gls{rrtfn}\cite{rrtfn}, \gls{rrtfnd} can generate routes between two points in an environment with obstacles that change over time.

\gls{rrt}* firstly builds routes through randomly adding nodes into an environment and then re-wiring the tree such that the cost from the origin node to any other node is minimized, with respect to all the nodes in the tree. This is referred to as the \texttt{Grow} function and forms the base of \gls{rrtfnd}. Operating identically to \gls{rrt}*, \gls{rrtfn} introduces the concept of \textit{fixed nodes} or, a maximum number of nodes ($M$) that can exist in the tree. The algorithm uses the \texttt{Grow} function until the number of nodes exceeds $M$ where \gls{rrtfn} employs a force removal algorithm. The algorithm, in short, randomly selects a childless node to remove from the tree so that the number of nodes in the tree remains no greater than $M$. There is a possibility that the last node added exists on the solution path. To avoid removing this newly added node it is excluded from the random selection of childless nodes.

\gls{rrtfnd}\cite{rrtfnd} improves \gls{rrtfn} for dynamic environments by utilizing existing information if the solution path is severed. Introduced are the \texttt{Reconnect} and \texttt{Regrow} functions. If the existing solution path is severed, \texttt{Reconnect} attempts to reconnect the solution tree directly to the original parent tree. If this connection cannot be made, then \texttt{Regrow} attempts to connect the parent tree to any node in the solution tree by rerunning the \texttt{Grow} function.

\subsubsection{RRT-Rope}
\gls{rrt}-Rope\cite{rrtrope} shortens existing paths using a technique similar to that of pulling a rope. This was added to, in addition to shortening a path, provide a smoother path, easier for an aircraft to follow as \gls{rrt}-Rope removes the frequent direction changes.

For each node in the solution tree, it looks at every subsequent node, starting with the furthest node first. It evaluates if a direct path has any collision with the obstacles in the environment. Providing no collision is found the two nodes will be directly connected moving intermediate nodes into a straight line between the two nodes.

\subsection{Route Generation using RRT*FND and RRT-Rope}
\subsubsection{Pre-processing}
To avoid errors over large distances when trying to use a constant interval with latitudes and longitudes, the airspace area was first converted to a local, flat Cartesian plane. This was performed using Equations \ref{eq:latitude}-\ref{eq:longitude} to obtain meters per degree of latitude/longitude at the latitude $\phi$\cite{Snyder1987,Bugayevskiy2013}.

\begin{equation}
    111132.92 - 559.82\cos{2\phi}+1.175\cos{4\phi}+0.0023\cos{6\phi}
    \label{eq:latitude}
\end{equation}
\begin{equation}
    111412.84\cos{\phi} - 93.5\cos{3\phi}+0.118\cos{5\phi}
    \label{eq:longitude}
\end{equation}

\subsubsection{Building Routes}
In this work we focus on the deconfliction against static \glspl{nfz} and existing \glspl{ov} in the airspace. Described in \cite{9298470} as ``\textit{keep-out} geofences'', the \glspl{nfz} are regions of airspace that both the route and accompanying \glspl{ov} must never overlap with. Unlike \glspl{nfz}, \glspl{ov} only exist for a finite duration ($t_d$) and so, from the perspective of the \gls{rrtfnd} algorithm, appear as dynamic obstacles appearing and disappearing with the progression of time. It should be noted that in this implementation we specifically deconflict \gls{ov} from the static \glspl{nfz} and focus on route-\gls{ov} for \glspl{ov} belonging to other contracts.

\begin{figure}[ht]
    \centering
    \begin{subfigure}[b]{\linewidth}
        \centering
        \includegraphics[width=0.8\linewidth]{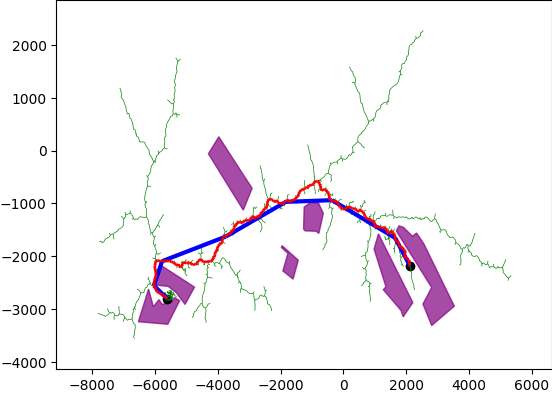}
        \caption{Optimization Factor: 50}
    \end{subfigure}
    \begin{subfigure}[b]{\linewidth}
        \centering
        \includegraphics[width=0.8\linewidth]{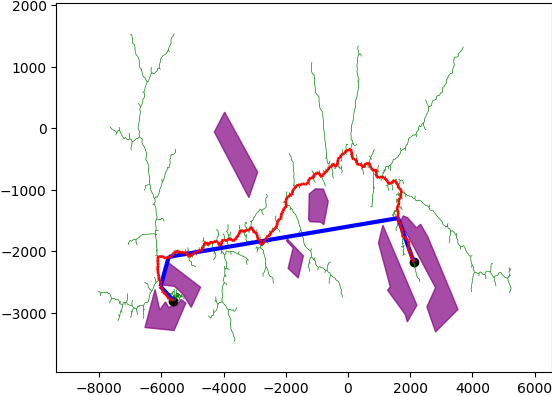}
        \caption{Optimization Factor: 10000}
    \end{subfigure}
    \caption{The \gls{rrtfnd} algorithm generating a valid route (red). The route has been further optimised (blue) to decrease the overall distance while avoiding conflicts, this is done based on the \gls{rrt}-Rope algorithm.}
    \label{fig:rrtfnd}
\end{figure}

As expected, we run the \gls{rrtfnd} until a valid candidate solution is generated and then optimised for distance using \gls{rrt}-Rope method. To improve computational efficiency, an optimisation factor is applied to the \gls{rrt}-Rope algorithm defining the maximum number of nodes to consider during the optimisation, this is demonstrated in Figure \ref{fig:rrtfnd}.

This candidate solution then checks for conflicts against existing \glspl{ov} by first checking for intersections between the route and \glspl{ov} and then estimating the arrival time at the intersection with a user-defined departure time and either a provided cruise speed or upper and lower bound for the speed. If a conflict is found, i.e. a time aligned intersection exists with an existing \gls{ov}, the system then checks the probability distribution of aircraft within the \gls{ov} using Equation \ref{eq: prob}. If the probability lies above a safety threshold, the candidate solution route is severed and the \texttt{Regrow} and \texttt{Reconnect} functions are used to repair the break with the addition of the \gls{ov} which the conflict was with appearing as a static obstacle.

Once a candidate solution is generated the \gls{ov} generation begins. After each \gls{ov} has been generated the aircraft positions, boundaries and associated times are compared with the \glspl{nfz} for violations. If a violation is found, violating nodes are removed from the graph and the \texttt{Reconnect} and \texttt{Regrow} functions of \gls{rrtfnd} are again used to dynamically reconnect the graph. 

The cycle is run until either a valid route is found or a timeout is reached, in which the system has not found a valid solution in a given time and so needs to be re-executed under different initialising parameters.

\section{EXPERIMENTS}
\label{sec:experiments}

\subsection{Evaluation of OV Generation}
To evaluate \gls{ov} generation three predefined routes were used. \glspl{ov} were then generated for these routes using the method outlined in Section \ref{sec: contracts and ovs}. Two experiments were designed to evaluate the performance of this approach: The first to test the natural point inclusion of \glspl{ov} for flights using autopilot; The second to test the sensitivity to initial conditions and the related size of the generated \glspl{ov}.

\subsubsection{Point inclusion}
In this experiment simulations were run in BlueSky with 1000 aircraft following the routes and their state data being recorded every second. The recorded data was then compared, both spatially and temporally, with the generated \gls{4d} \glspl{ov} recording the distribution of points that existed within a valid \gls{ov} given their relative times.

The three test scenarios generated tested the inclusion of recorded points within the \glspl{ov}. Beyond the autopilot logic implemented in BlueSky the aircraft were not controlled to specifically remain within the \glspl{ov}. Three scenarios were created for the experiment testing: (1) relatively normal operating over a $\approx15$-minute route; (2) a significantly shorter circular route $\approx8$-minute flight time; (3) a complex route with frequent speed and altitude changes and aggressive turns ($>120^{\circ}$), this route was equivalent roughly to an approximate $25$-minute flight.

The results can be seen in Table \ref{tab:OV Results}, and demonstrate high performance within the first two scenarios. The poorer performance in Scenario 3, can be explained through the frequency and aggressiveness of speed and heading changes and the time it takes an aircraft to perform the manoeuvre. 

\begin{table}[ht]
\caption{\gls{ov} point inclusion results}
\label{tab:OV Results}
\begin{center}
\begin{tabular}{|c||c|c|}
\hline
Scenario   & \begin{tabular}[c]{@{}c@{}}Total \\ recorded points\end{tabular} & \begin{tabular}[c]{@{}c@{}}\% of points \\ in valid OV\end{tabular} \\ \hline
1 & 732102                                                           & 98.52\%                                                             \\ \hline
2 & 553717                                                           & 92.19\%                                                             \\ \hline
3 & 1176679                                                          & 79.55\%                                                             \\ \hline
\end{tabular}
\end{center}
\end{table}

\subsubsection{Sensitivity of \gls{ov} generation to initial conditions}
In this experiment the initial conditions of the aircraft were altered to provide larger uncertainty. The internal volume of the \glspl{ov}, measured in kilometres$^2$ ($km^2$), was then recorded, along with the number of \glspl{ov} generated. For this experiment we used the same three test routes as mentioned previously. We measured different distributions of initial speed, position and altitude. 

Table \ref{tab:OV sensitivity speed} shows the results with variation in the initial speed. The initial position remained constant, and the altitude was between $[5m,10m]$. In Case 1 speeds were generated randomly between $20m/s$ and $24m/s$ and in Case 2 between $16m/s$ and $26m/s$. The results show that, with a larger range of possible initial speeds, the solution generates more \glspl{ov} and of a larger size. The increase in the number of \glspl{ov} is likely a result of a larger distribution over the points used when initialising aircraft states from the normal distribution described in Section \ref{sec: contracts and ovs}.

\begin{table}[ht]
\centering
\caption{OV sensitivity to variation in initial speeds}
\label{tab:OV sensitivity speed}
\begin{tabular}{|c||cc|cc|}
\hline
Scenario   & \multicolumn{2}{c|}{\begin{tabular}[c]{@{}c@{}}Speed \\ $[20m/s, 24m/s]$\end{tabular}}                                       & \multicolumn{2}{c|}{\begin{tabular}[c]{@{}c@{}}Speed \\ $[16m/s, 26m/s]$\end{tabular}}                                        \\ \cline{2-5} 
           & \multicolumn{1}{c|}{\begin{tabular}[c]{@{}c@{}}Number \\ of OVs\end{tabular}} & \begin{tabular}[c]{@{}c@{}}Mean \\ Area\end{tabular} & \multicolumn{1}{c|}{\begin{tabular}[c]{@{}c@{}}Number \\ of OVs\end{tabular}} & \begin{tabular}[c]{@{}c@{}}Mean \\ Area\end{tabular} \\ \hline
1 & \multicolumn{1}{c|}{18}                                                       & 0.44                                              & \multicolumn{1}{c|}{18}                                                       & 1.3                                            \\ \hline
2 & \multicolumn{1}{c|}{12}                                                       & 0.1                                               & \multicolumn{1}{c|}{13}                                                      & 0.1                                               \\ \hline
3 & \multicolumn{1}{c|}{28}                                                      & 0.03                                                  & \multicolumn{1}{c|}{29}                                                      & 0.13                                                \\ \hline
\end{tabular}
\end{table}

In the next experiment, to test sensitivity to initial conditions, changes to the initial positional variation were implemented. Initial starting speed was fixed to $18m/s$ but, on the same three scenarios, the area aircraft could be generated in initially was altered. In the first test, from the initial starting points, a variation of $\pm 10m$ was introduced, $\pm 50m$ for the second and $\pm 100m$ for the third in any direction. As is expected generated \glspl{ov} are larger if there is a greater variation in initial starting positions, these results are demonstrated in Table \ref{tab:OV sensitivity position}.

\begin{table}[ht]
\centering
\caption{OV sensitivity to variation in starting position}
\label{tab:OV sensitivity position}
\begin{tabular}{|c||cc|cc|cc|}
\hline
Scenario   & \multicolumn{2}{c|}{\begin{tabular}[c]{@{}c@{}}Starting Area \\ $\pm 10m$\end{tabular}}                                       & \multicolumn{2}{c|}{\begin{tabular}[c]{@{}c@{}}Starting Area \\ $\pm 50m$\end{tabular}} & \multicolumn{2}{c|}{\begin{tabular}[c]{@{}c@{}}Starting Area \\ $\pm 100m$\end{tabular}}                                        \\ \cline{2-7} 
           & \multicolumn{1}{c|}{\begin{tabular}[c]{@{}c@{}}Number \\ of OVs\end{tabular}} & \begin{tabular}[c]{@{}c@{}}Mean \\ Area\end{tabular} & \multicolumn{1}{c|}{\begin{tabular}[c]{@{}c@{}}Number \\ of OVs\end{tabular}} & \begin{tabular}[c]{@{}c@{}}Mean \\ Area\end{tabular}& \multicolumn{1}{c|}{\begin{tabular}[c]{@{}c@{}}Number \\ of OVs\end{tabular}} & \begin{tabular}[c]{@{}c@{}}Mean \\ Area\end{tabular}\\ \hline
1 & \multicolumn{1}{c|}{18}                                                       & 0.142                                             & \multicolumn{1}{c|}{18}                                                       & 0.203 
& \multicolumn{1}{c|}{18}                                                       & 0.316                                           \\ \hline
2 & \multicolumn{1}{c|}{12}                                                       & 0.164                                               & \multicolumn{1}{c|}{12}                                                      & 0.142
& \multicolumn{1}{c|}{12}                                                       & 0.313 \\ \hline
3 & \multicolumn{1}{c|}{28}                                                      & 0.032                                                  & \multicolumn{1}{c|}{28}                                                      & 0.038 
& \multicolumn{1}{c|}{28}                                                       & 0.046 
\\ \hline
\end{tabular}
\end{table}

\subsection{Evaluating Route Generation}
\begin{figure}[ht]
    \centering
    \includegraphics[width=0.65\linewidth]{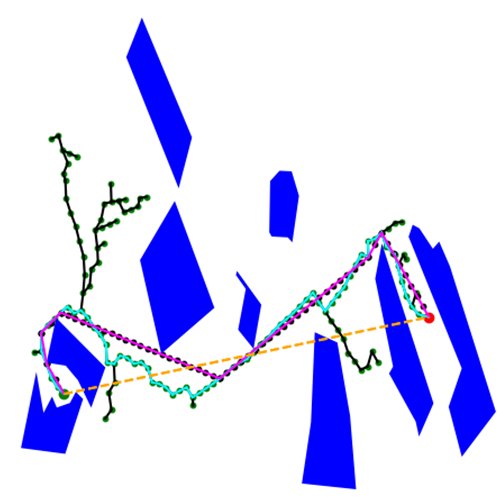}
    \caption{Example routes generated with the cyan line being the original candidate route ($10113m$), the magenta line is the optimised route ($9443$), and the dashed orange line represents the direct route ($7761m$).}
    \label{fig:shortest_routes}
\end{figure}

To evaluate the performance of candidate route generation the generation time and overall route length were the concerning factors. Figure \ref{fig:shortest_routes} shows the example environment that the routes were generated. The dark blue regions represent static \glspl{nfz}. Additionally, the candidate route is denoted in cyan, optimised route is coloured magenta and the dashed orange line represents the direct distance, which is $7761m$.

Two experiments were run to evaluate performance: (1) evaluates the run-time of the route generation, this includes time to optimise the route as well; (2) evaluates the optimisation performance, examining how significantly the route can be shortened, using the maximum optimisation factor.

\subsubsection{Route generation time evaluation}
In this experiment the route generation was run with varying step sizes, maximum distance a node can be from any other node, to evaluate run time. Table \ref{tab:gen_times} shows the results of this experiment. Intuitively, the generation time increases with the resolution of the step size. Pertaining to the \textit{maximum nodes in graph} the $50m$ resolution step size was unable to generate a valid route with a maximum of 150 nodes in the graph, for this test the maximum number of nodes was increased to 300.

\begin{table}[ht]
\caption{Time to generate route}
\begin{center}
    \begin{tabular}{|c||c|c|}
    \hline
       Resolution    & Time (s) & \begin{tabular}[c]{@{}c@{}}Maximum Nodes \\ in Graph\end{tabular} \\ \hline
50 meters  & 108.82   & 300                                                               \\ \hline
100 meters & 19.90    & 150                                                               \\ \hline
150 meters & 6.84     & 150                                                               \\ \hline
200 meters & 1.85     & 150                                                               \\ \hline
\end{tabular}
\end{center}
\label{tab:gen_times}
\end{table}

\subsubsection{Route optimisation evaluation}
To evaluate the performance of the \gls{rrt}-Rope based optimisation, the lengths of the routes were compared. Again, the resolution of the step size was also considered. Table \ref{tab:optimisation eval} shows there appears to be little correlation between the resolution of the step size and the difference in lengths. Each resolution was able to decrease the overall route length by more than $1000m$ where routes with step sizes $50m$ and $150m$ decreased the route length by more than $1500m$.

A significant improvement, demonstrated in Figure \ref{fig:rrtfnd} and Figure \ref{fig:shortest_routes}, is that the optimisation significantly decreases the complexity of the route. While the \gls{rrtfnd} candidate routes are valid they consist of many sharp or aggressive turns. In flight this could cause significant stress on the vehicle or, increase flight duration significantly. By reducing the path to straight line segments, the optimisation not only reduces the overall length but decreases the complexity of the route significantly.

\begin{table}[ht]
\caption{Comparison of candidate and optimised route lengths}
\begin{center}
\begin{tabular}{|c||c|c|c|}
\hline
     Resolution      & \begin{tabular}[c]{@{}c@{}}Candidate Length\\ ($m$)\end{tabular} & \begin{tabular}[c]{@{}c@{}}Optimised Length\\ ($m$)\end{tabular} & \begin{tabular}[c]{@{}c@{}}$\Delta$ Length\\ ($m$)\end{tabular} \\ \hline
50 meters  & 11151.01                                                         & 9369.89                                                          & 1781.12                                                         \\ \hline
100 meters & 10975.74                                                         & 9525.31                                                          & 1450.43                                                         \\ \hline
150 meters & 11335.55                                                         & 9658.40                                                          & 1677.16                                                         \\ \hline
200 meters & 11096.91                                                         & 9836.05                                                          & 1260.86                                                         \\ \hline
\end{tabular}
\label{tab:optimisation eval}
\end{center}
\end{table}

\section{DISCUSSIONS}
\label{sec: discussion}

\subsection{Generation of Operational Volumes}

In the approach presented in Section \ref{sec: contracts and ovs}, reachability analysis was performed using the DryVR\cite{dryvr} framework to generate reach tubes. These reach tubes provided the \gls{3d} space of the \glspl{ov}. The reachability analysis performed in DryVR leverages a learned \textit{Discrepancy Function} from the sensitivities obtained in the simulation traces. As such, with enough uncertainty or noise in the data, this approach begins to suffer from the \textit{wrapping effect}\cite{Neumaier1993}, resulting in over conservative, non-descriptive \glspl{ov} that grow in size exponentially. The results in Table \ref{tab:OV sensitivity speed} and Table \ref{tab:OV sensitivity position} show the \glspl{ov} growing in size with more introduced uncertainty. While uncertainty was presented manually, one source of natural uncertainty can be the weather or, more commonly, wind. Including predicted winds into the simulation when generating the \glspl{ov} will improve the overall model and provide more accurate simulation data to generate the \glspl{ov}.

An improved method may look at using an analytical approach to reach tube generation. A Lagrangian approach, such as in \cite{ctrl,Gruenbacher_2020,gotube} has shown to provide tighter bounded reach tubes and claim to not suffer from the \textit{wrapping effect}\cite{Gruenbacher_2020}. While the DryVR-based approach did not necessarily suffer from the \textit{wrapping effect} the test routes presented were relatively small, larger or more complex routes may result in additional problems.

\subsection{Route Generation and Deconfliction}
The \gls{rrt} based algorithm used (\gls{rrtfnd}) is not known to generate optimal shortest routes. While the approach is suited to a dynamic environment there is limited control on the optimality of the resultant route. As a result, improving this algorithm could be achieved through implementing stricter constraints in where valid nodes can be placed or using an algorithm such as D* or D* Lite\cite{351061,dstarlite}. Both algorithms provide the dynamic requirement while yielding optimal, shortest-path trajectories. D* has also been proven to be adaptable to route planning for aircraft in\cite{7566733}, supporting the adoption further.

Additionally, refinements to either the aircraft positional probability or collision probability within a cost function could also improve the overall performance of the system. Presently, the approach only assigns probabilities to within a 10 meter $\times$ 10-meter region of airspace. This presents the problem of routes potentially approaching too closely to potential conflict areas. To improve this, extending regions of high values such that the values diminish in a gradient could discourage the path finding algorithm from approaching too closely to potential conflict regions.

\section{CONCLUSION}
\label{sec:conclusion}

In this paper we presented a framework for generating and deconflicting routes in conjunction with the generation of \gls{4d} \glspl{ov} to address airspace and operational uncertainties. We used a reachability analysis method built on the DryVR\cite{dryvr} framework to generate \glspl{ov} given simulation data obtained through BlueSky\cite{bluesky}. An approach to route generation and deconfliction was given using an adaptation of the \gls{rrt} algorithm: \gls{rrtfnd}. This was then, optimized for shortened distance using the concept of rope-pulling described in \cite{rrtrope}. Results demonstrated the ability to generate candidate routes, deconflict them with static \glspl{nfz} and generate a contract of \glspl{ov} based on the reachability analysis adapted from DryVR\cite{dryvr}. We also provide an oversight into a potential adaptation of this method that could remove the potential of \gls{ov} generation to suffer from the wrapping effect by adapting the Lagrangian approach to reachability analysis in \cite{gotube}.

The approach presented in this paper, is intended to be performed pre-flight, however, due to the choice in dynamic path planning algorithm, (\gls{rrtfnd}), there is prospect for an in-flight, re-planning approach that largely utilises the same foundation. This is further reinforced by the path planning and \gls{ov} generation occurring simultaneously. Additionally, proposed future work will improve both the route generation approach as well as \gls{ov} generation, and introduce the strategic functionality of deconflicting against existing contracts of \glspl{ov} accurately and efficiently.

Furthermore, we intend to develop a tactical, in-flight layer be added to ensure safe separation of en route aircraft. We intend to introduce such a tactical system, leveraging the constraints imposed by the aforementioned \glspl{ov}. That is, a tactical separation assurance component will provide instructions to aircraft to ensure a minimum separation is maintained by aircraft, both laterally and vertically, and ensure the aircraft remain within their contracted \glspl{ov}.

\bibliographystyle{IEEEtran}
\bibliography{bibliography}

\end{document}